\documentstyle[12pt,graphicx]{article}
\title{Classical and quantum regimes of the superfluid
turbulence. }
\author{G.E. Volovik\\
Low Temperature Laboratory,
Helsinki University of Technology\\
P.O.Box 2200, FIN-02015 HUT, Finland\\
and\\
L.D. Landau Institute for Theoretical Physics RAS \\
 Kosygina 2, 117940
Moscow, Russia }
\begin{document}
\maketitle
\begin{abstract}
{We argue that turbulence in superfluids is governed by two
dimensionless parameters. One of them is the intrinsic parameter
$q$ which characterizes the friction forces acting on a vortex moving
with respect to the heat bath, with $q^{-1}$ playing the
same role as the  Reynolds number ${\rm Re}=UR/\nu$ in classical
hydrodynamics. It marks the transition between the "laminar" and turbulent
regimes of vortex dynamics. The developed turbulence described by
Kolmogorov cascade occurs when ${\rm Re}\gg 1$ in classical
hydrodynamics, and $q\ll 1$ in the superfluid hydrodynamics. Another
parameter of the superfluid turbulence is the superfluid Reynolds number
${\rm Re}_s=UR/\kappa$, which contains the circulation quantum $\kappa$
characterizing quantized vorticity in superfluids. This parameter
may regulate the crossover or transition between two classes of
superfluid turbulence: (i) the classical regime of Kolmogorov cascade
where vortices are locally polarized and the quantization of vorticity is
not important; and (ii) the quantum Vinen turbulence whose properties are
determined by the quantization of vorticity. The phase diagram of the
dynamical vortex states is suggested.}
\end{abstract}


\section{Introduction}

The hydrodynamics of superfluid liquid exhibits new features
with respect to conventional classical hydrodynamics, which are
important when the turbulence in superfluids is considered
\cite{VinenNiemela}. 

(i) The superfluid liquid consists of two mutually penetrating components
-- the frictionless superfluid component and the viscous normal 
component. That is why different types of turbulent motion are
possible depending on whether the normal and superfluid components move
together or separately. Here we are interested in the most simple case
when the dynamics of the normal component can be neglected. This occurs,
for example, in superfluid phases of $^3$He where the normal component is
so viscous that it is practically clamped by the container walls. Its role
is to provide the preferred reference frame, where the normal component
and thus the heat bath are at rest. The turbulence in the superfluid
component with the normal component at rest is referred to as
the superfluid turbulence.

(ii) The important feature of the  superfluid turbulence is that the
vorticity of the superfluid component is quantized in terms of
the elementary circulation quantum $\kappa$. So the superfluid
turbulence is the chaotic motion of well determined and well
separated vortex filaments \cite{VinenNiemela}. Using this we can
simulate the main ingredient of the classical turbulence -- the chaotic
dynamics of the vortex degrees of freedom of the liquid. 

(iii) The further simplification comes from the fact that the dissipation
of the vortex motion is not due to the viscosity term in the Navier-Stokes
equation which is proportional to the velocity gradients $\nabla^2{\bf v}$
in classical liquid, but due to the friction force acting on the vortex
when it moves with respect to the heat bath (the normal component). This
force is proportional to velocity of the vortex, and thus the
complications resulting from the $\nabla^2{\bf v}$ term are avoided. 

Here we discuss how these new features could influence the superfluid
turbulence. 

\section{Coarse-grained hydrodynamic equation}

The coarse-grained hydrodynamic equation for the superfluid vorticity  is
obtained from the Euler equation for the superfluid velocity  ${\bf
v}\equiv {\bf v}_{\rm s}$  after averaging over the vortex lines  
(see the review paper \cite{Sonin}). Instead of the Navier-Stokes
equation with
$\nabla^2{\bf v}$ term one has
\begin{eqnarray}
\frac{\partial {\bf v} }{ \partial t}+ \nabla\mu=  {\bf
v} \times
 \vec{\bf\omega}~-
\label{SuperfluidHydrodynamics1}
\\
 -\alpha'({\bf
v} -{\bf v}_{\rm
n})\times
 \vec{\bf\omega}+  \alpha~\hat{\bf\omega}
\times(\vec{\bf\omega}
\times({\bf
v} -{\bf v}_{\rm
n}) ) ~.
\label{SuperfluidHydrodynamics2}
\end{eqnarray} 
Here ${\bf v}_{\rm n}$  is the velocity of the normal component;
$\vec{\bf\omega}=\nabla\times {\bf v}$ is the superfluid vorticity; 
$\hat\omega=\vec{\bf\omega}/\omega$; and dimensionless parameters
$\alpha'$ and $\alpha$ come from the reactive and dissipative forces
acting on a vortex when it moves with respect to the normal component.
These parameters are very similar to the Hall resistivity
$\rho_{xy}$ and $\rho_{xx}$ in the Hall effect. For vortices in fermionic
systems (superfluid $^3$He and superconductors) they were calculated by
Kopnin
\cite{KopninBook}, and measured in $^3$He-B in the broad temperature
range by Bevan et. al.
\cite{Bevan} (see also \cite{VolovikBook}, where these parameters are
discussed in terms of the chiral anomaly).

The terms in
Eq.(\ref{SuperfluidHydrodynamics1}) are invariant with respect to the
transformation ${\bf v}\rightarrow {\bf v}({\bf r}-{\bf u}t) +{\bf u}$ as
in classical hydrodynamics. However, the terms in
Eq.(\ref{SuperfluidHydrodynamics2}) are not invariant under this
transformation, since there is the
preferred reference frame in which the normal component is at rest. 
They are invariant under the full Galilean transformation
when the normal component is also involved: ${\bf v}\rightarrow {\bf
v}({\bf r}-{\bf u}t) +{\bf u}$ and 
${\bf v}_{\rm n}\rightarrow {\bf v}_{\rm n} +{\bf u}$.

Further we shall work in the frame where   
${\bf v}_{\rm n}=0$, but we must remember that this frame is unique. 
In this frame the equation for superfluid hydrodynamics is simplified:
\begin{equation}
\frac{\partial {\bf v} }{ \partial t}+ \nabla\mu= (1-\alpha'){\bf
v} \times
 \vec{\bf\omega}+  \alpha~\hat{\bf\omega} \times(\vec{\bf\omega}
\times{\bf v} ) ~.
\label{SuperfluidHydrodynamics}
\end{equation} 
After rescaling the time,
$\tilde t=(1-\alpha')t$, one obtains equation which depends on a single
parameter $q=\alpha/(1-\alpha')$:
\begin{equation}
\frac{\partial {\bf v} }{ \partial \tilde t}+ \nabla \tilde\mu=
 {\bf v} \times
 \vec{\bf\omega}+  q~\hat{\bf\omega} \times(\vec{\bf\omega}
\times{\bf v} ) ~.
\label{SuperfluidHydrodynamics3}
\end{equation} 
Now the first three terms together are the same as inertial terms in
classical hydrodynamics. They satisfy the modified Galilean
invariance (in fact the transformation below changes the chemical
potential, but this does not influence the vortex degrees of freedom
which are important for the  phenomenon of turbulence): 
\begin{equation}
 {\bf v}( \tilde t,{\bf r}) \rightarrow {\bf v}( \tilde t,{\bf r}-{\bf
u}\tilde t)+{\bf u} ~.
\label{GalileanInvariance}
\end{equation}
 On the contrary the
dissipative last term with the factor
$q$ in Eq.(\ref{SuperfluidHydrodynamics3}) is not invariant under this
transformation.  This is in
contrast to the conventional liquid where the whole Navier-Stokes equation
which contains viscosity
\begin{equation}
\frac{\partial {\bf v}}{ \partial t}+ \nabla\mu=  {\bf v}\times
 \vec{\bf\omega}+  \nu\nabla^2 {\bf v}
~,
\label{NormalHydrodynamics}
\end{equation} 
is Galilean invariant, and where there is
no preferred reference frame. 

 Such a difference between the dissipative last
terms in Eqs. (\ref{NormalHydrodynamics}) and
(\ref{SuperfluidHydrodynamics3}) is very important: 

(1) The role of the Reynolds number, which characterizes the ratio of
inertial and dissipative terms in hydrodynamic equations,  in the
superfluid turbulence  is played by the intrinsic parameter
$1/q$. This parameter does not depend on the characteristic velocity $U$
and size
$R$ of the large-scale flow as distinct from the conventional
Reynolds number ${\rm
Re}=RU/\nu$ in classical viscous hydrodynamics. That is why the
turbulent regime occurs only at
$1/q>1$ even if vortices are injected to the superfluid which moves with
large velocity $U$. This rather unexpected result was obtained in recent
experiments with superfluid $^3$He-B \cite{Finne}. 

(2) In the conventional
turbulence the large-scale velocity $U$ is always understood as the
largest characteristic velocity difference in the inhomogeneous
flow of classical
liquid \cite{McComb}.
In the two-fluid system the velocity 
$U$ is  the large-scale velocity of
superfluid component with respect to the normal component, and this
velocity (the so-called counterflow velocity) can be completely
homogeneous. 

(3) As a result, as distinct from the classical hydrodynamics,  the energy
dissipation which is produced by the last term in
Eq.(\ref{SuperfluidHydrodynamics3}) depends explicitly on
$U$:
\begin{equation}
 \epsilon =-\dot E=-\langle{\bf v}\cdot \frac{\partial {\bf v} }{ \partial
\tilde t}\rangle=  -q \langle {\bf v}\cdot(\hat{\bf\omega}
\times(\vec{\bf\omega}
\times{\bf v} ))\rangle \sim  q \omega
U^2 ~.
\label{EnergyDissipation}
\end{equation}

(4)  The onset of the superfluid turbulence was studied in
Ref. \cite{Kopnin}, where the model was developed which demonstrated that
the initial avalanche-like multiplication of vortices leading to
turbulence occurs when $q< 1$ in agreement with experiment
\cite{Finne}.
 The existence of two regimes in the initial development of vorticity is
also supported by earlier simulations
by Schwarz who noted that when $\alpha$ (or $q$) is decreased the
crossover from a regime of isolated phase slips to a phase-slip cascades
and then to the fully developed vortex turbulence occurs
\cite{Schwarz}.
One can expect that the well developed turbulence occurs when
$q\ll 1$, and here we shall discuss this extreme limit. In
$^3$He-B the condition $q\ll 1$ can be ralized at temperatures well
below
$0.6T_c$
\cite{Finne}. However, we do not consider a very low $T$ where instead
of the mutual friction the other mechanisms of dissipation take place
such as excitation of Kelvin waves
\cite{KelvinWaveCascade} and vortex reconnection
\cite{TsubotaArakiNemirovskii}. The latter leads to formation of cusps
and kinks on the vortex filaments whose fast dynamics creates the burst
of different types of excitations in quantum liquids: phonons, rotons,
Kelvin waves and fermionic quasiparticles. The burst of gravitational
waves from cusps and kinks of cosmic strings was theoretically
investigated by the cosmological community (see e.g. 
\cite{DamourVilenkin}), and the obtained results are very important
for the superfluid turbulence at a very low temperature.

(5) We expect that even at  $q\ll 1$ two
different states of turbulence are possible, with the crossover (or
transition) between them being determined by $q$ and by another
dimensionless parameter
${\rm Re}_s=UR/\kappa$, where $\kappa$ is the circulation around the
quantum vortex. The coarse-grained hydrodynamic equation
(\ref{SuperfluidHydrodynamics3}) is in fact valid only in the limit 
${\rm Re}_s\gg 1$, since the latter means that the characteristic
circulation of the velocity $\Gamma=UR$ of the large-scale
flow substantially exceeds the circulation quantum $\kappa$, and thus
there are many vortices in the turbulent flow. When ${\rm Re}_s$ 
decreases the quantum nature of vortices becomes more pronounced, and we
proceed from the type of the classical turbulence which is probably
described by the Komogorov cascade, to the quantum regime which is
probably described by the Vinen equations for the average vortex dynamics
\cite{Vinen}.

Let us consider the possibility of the Kolmogorov
state of the superfluid turbulence.

\section{Kolmogorov cascade}

 In classical turbulence, the large Reynolds number $Re=UR/\nu\gg 1$
leads to the well separated length scales or wave numbers. As a result the
Kolmogorov-Richardson cascade takes place in which the energy flows
from small wave numbers $k_{min}\sim 1/R$ (large rings of size $R$ of the
container) to high wave number $k_0=1/r_0$ where the dissipation occurs. 
In the same manner in our case of the superfluid turbulence the
necessary condition for the  Kolmogorov cascade is the big
ratio of the inertial and dissipative terms in
Eq.(\ref{SuperfluidHydrodynamics3}), i.e.   
$1/q\gg 1$. 

In the  Kolmogorov-Richardson cascade,
at arbitrary length scale  $r$ the energy transfer rate to
the smaller scale, say $r/2$, is $\epsilon=  E_r/ t_r$, where $E_r=v_r^2$
is the kinetic energy at this scale, and $t_r=r/v_r$ is the
characteristic time. The energy transfer from scale to scale must be
the same for all scales, as a result one has
\begin{equation}
 \epsilon= \frac{E_r}{ t_r}=\frac{v_r^3}{ r}={\rm
constant}=\frac{U^3}{ R}~.  
\label{Micriscopic1}
\end{equation}
From this equation it follows that
\begin{equation}
v_r=  \epsilon^{1/3}~ r^{1/3}~.  
\label{Micriscopic2}
\end{equation}
This must be valid both in classical and superfluid
liquids \cite{TsubotaKasamatsuAraki}. What is different is the
parameter
$\epsilon$: it is determined by the dissipation mechanism which is
different in two liquids.

 From Eq.(\ref{EnergyDissipation}) with
$\omega_r=v_r/r$ it follows that as in the classical turbulence the main
dissipation occurs at the smallest possible scales, but the structure of
$\epsilon$ is now different. Instead of $\epsilon=\nu  v_{r0}/ r_0^2$ in
classical liquids, we have now
\begin{equation}
 \epsilon \sim  q \omega_{r0} U^2 \sim qU^2 \frac{v_{r0}}{ r_0}=
 qU^2 \epsilon^{1/3}~ r_0^{-2/3}~.  
\label{EnergyDissipation2}
\end{equation}
Since $\epsilon=U^3/R$ one obtains from Eq.(\ref{EnergyDissipation2})
that the scale $r_0$ at which the main dissipation occurs and the
characteristic velocity $v_{r_0}$ at this scale are
\begin{equation}
   r_0\sim  q^{3/2}R ~~,~~v_{r0}\sim
q^{1/2}U~.  
\label{MinimalScale}
\end{equation}
This consideration is valid when $r_0\ll R$ and $v_{r0}\ll U$, which
means that $1/q\gg 1$ is the condition for the Kolmogorov cascade. In
classical liquids the corresponding condition for the well developed
turbulence is
$Re\gg 1$. In both cases these conditions ensure that the kinetic
terms in the hydrodynamic equations are much larger than the
dissipative terms. In the same manner as in classical liquids the
condition for the stability of the turbulent flow is $Re>1$, one may
suggest that the condition for the stability of
the discussed turbulent flow is $1/q>1$. This is
supported by observations in $^3$He-B where it was demonstrated that
at high velocity $U$ but at
$q>1$ the turbulence is not developed even after vortices were
introduced into the flow
\cite{Finne}.

As in the Kolmogorov cascade for the classical liquid, in the
Kolmogorov cascade of superfluid turbulence the dissipation is
concentrated at small scales, 
\begin{equation}
 \epsilon \sim  q   U^2 \int_{r_0}^R\frac{dr}{ r}\frac {v_{r}}{ r}\sim
qU^2 \frac{v_{r0}}{ r_0}~,  
\label{EnergyDissipation3}
\end{equation}
while the kinetic
energy is concentrated at large scale of container size:
\begin{equation}
 E=\int_{r_0}^R\frac{dr}{ r} v_r^2=\int_{r_0}^R\frac{dr}{ r} (\epsilon
r)^{2/3}= (\epsilon R)^{2/3} =U^2~.  
\label{KineticEnergyR}
\end{equation}

The dispersion of the turbulent energy in the momentum space is the same
as in classical liquid
\begin{eqnarray}
 E= \int_{r_0}^R\frac{dr}{ r} (\epsilon
r)^{2/3}=\int_{k_0}^{1/R} \frac{dk}{ k}  \frac{\epsilon^{2/3}}{ k^{2/3}}
=\int_{k_0}^{1/R} dk E(k),
\nonumber
\\
  E(k)=\epsilon^{2/3} k^{-5/3}.  
\label{KineticEnergyK2}
\end{eqnarray}
As distinct from the classical liquid where $k_0$ is
determined by viscosity, in the superfluid turbulence the cut-off
$k_0$ is determined by mutual friction papameter $q$:
$k_0=1/r_0=R^{-1}q^{-3/2}$.

\section{Crossover to Vinen quantum turbulence}

At a
very small $q$ the quantization of circulation becomes important.
The condition of the above consideration is that the
relevant circulation can be considered as continuous, i.e. the
circulation at the scale $r_0$ is  larger than the circulation
quantum:
$v_{r0}r_0> 
\kappa$. This gives 
\begin{equation}
 v_{r0} r_0 =q^2 U R=q^2 \kappa {\rm Re}_s  > \kappa
~~,~~{\rm Re}_s=\frac{UR}{\kappa}~,
\label{PhaseBoundary1}
\end{equation}
i.e. the constraint for the application of the Kolmogorov cascade is
\begin{equation}
{\rm Re}_s > \frac{1}{  q^2}\gg 1  ~ .  
\label{PhaseBoundary2}
\end{equation}
Another requirement is that the characteristic scale $r_0$ must be much
larger that the intervortex distance $l$. The latter is obtained from the
vortex density in the Kolmogorov state
$n_{\rm K}=l^{-2}=\omega_{r0}/\kappa= v_{r0}/(r_0  
\kappa)$. The condition $l\ll r_0$ leads again to the equation
$v_{r0}r_0> 
\kappa$ and thus to the criterion (\ref{PhaseBoundary2}).

Note that here for the first time the `superfluid Reynolds number'
${\rm Re}_s$ appeared, which contains the circulation quantum. Thus the
superfluid Reynolds number is responsible for the crossover
or transition from the classical superfluid turbulence, where the
quantized vortices are locally aligned (polarized), and thus the
quantization is not important, to the quantum  
turbulence developed by Vinen. 

We can now consider the approach to the crossover from the quantum
regime -- the Vinen state which probably occurs when ${\rm Re}_s q^2<1$.
According to Vinen \cite{Vinen} the characteristic length scale, the
distance between the vortices or the size of the characteristic vortex
loops, is determined  by the circulation quantum and the counterflow
velocity,
$l=\lambda
\kappa /U$, where
$\lambda$ is the dimensionless intrinsic parameter, which probably
contains
$\alpha'$ and
$\alpha$. The vortex density in the Vinen state is
\begin{equation}
 n_{\rm V}=l^{-2}\sim \lambda^2     \frac{U^2}{\kappa^2}=     
\frac{\lambda^2}{ R^2}  {\rm Re}_s^2 ~,  
\label{VortexDensityVinen}
\end{equation}
It differs from the vortex density in the Kolmogorov state
\begin{equation}
 n_{\rm K}=\frac{v_{r0}}{\kappa r_0}  \sim   \frac {U }{ q\kappa R}
= \frac{1}{ R^2} \frac{{\rm Re}_s}{ q}~~,  
\label{VortexDensity}
\end{equation}
which depends not only on the counterflow velocity $U$, but also on
the container size $R$.

\begin{figure}
  \centerline{\includegraphics[width=0.8\linewidth]{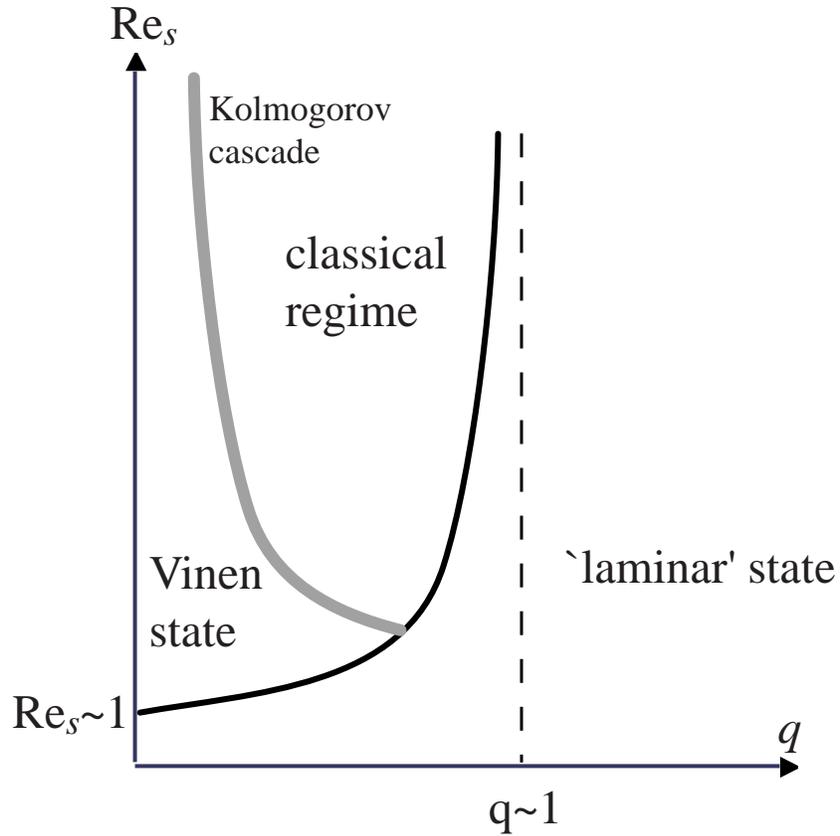}}
  \caption{Possible phase diagram of dynamical vortex states in ${\rm
Re}_s,q$ plane. At large flow velocity ${\rm Re}_s\gg 1$ the boundary
between the turbulent and `laminar' vortex flow approaches the vertical
axis
$q=q_0\sim 1$ as suggested by experiment \protect\cite{Finne}. The thick
line separates the developed turbulence of the classical type, which is
characterized by the Kolmogorov cascade at small
$q$, and the quantum turbulence of the Vinen type. }
  \label{PhaseDiagram}
\end{figure}

If the crossover between the classical and quantum regimes of the
turbulent states occurs at ${\rm Re}_s q^2=1$, the two equations
(\ref{VortexDensityVinen}) and (\ref{VortexDensity}) must match each other
in the crossover region. But this occurs only if
$\lambda^2= q$. If $\lambda^2\neq q$ there is a mismatch, and one may
expect that either the two states are separated by the first-order phase
transition, or there is an intermediate region where the superfluid
turbulence is described by two different microscopic scales such as
$r_0$ and
$l$. Based on the above consideration one may suggest the following phase
diagram of different regimes of collective vortex dynamics in Fig. 1. 

\section{Discussion}

The superfluid turbulence is the collective many-vortex phenomenon which 
can exist in different states. Each of the vortex states can be
characterized by its own correlation functions. For example, the states
can be characterized by the loop function 
\begin{equation}
g(C)=\left<e^{i(2\pi/\kappa)\oint_C {\bf v}\cdot d{\bf r}}
\right>~.
\label{LoopFunction}
\end{equation}
In the limit when the length $L$ of the loop $C$ is much larger than
the intervortex distance $l$ one may expect the general behavior $g(L)\sim
\exp(-(L/l)^{\gamma})$ where the exponent
$\gamma$ is different for different vortex states. 

One can expect the phase transitions between different states of
of collective vortex dynamics. One of such transitions which appeared to
be rather sharp has been observed between the `laminar' and `turbulent'
dynamics of vortices in superfluid $^3$He-B \cite{Finne}. It was found
that such transition was regulated by intrinsic velocity-independent
dimensionless parameter
$q=\alpha/(1-\alpha')$. However, it is not excluded that  both 
dimensionless parameters $\alpha$ and $\alpha'$ are important, and also it
is possible that only the initial stage of the formation of the
turbulent state is governed by these parameters \cite{Kopnin}.  Another
transition (or maybe crossover) is suggested here between the
quantum and classical regimes of the developed superfluid
turbulence, though there are arguments that the classical regime
can never be reached because the vortex stretching is missing in
the superfluid turbulence \cite{Kivotides}. 

In principle the parameters $\alpha$ and $\alpha'$
may depend on the type of the dynamical state, since they are obtained by
averaging of the forces acting on individual vortices. The
renormalization of these parameters $\alpha(L)$ and $\alpha'(L)$ when the
length scale $L$ is increasing
 may also play an important role in the identification of the turbulent
states, as in the case of the renormalization-group flow of similar
parameters in the quantum Hall effect (see e.g. \cite{Khmelnitskii}).

I thank V.B. Eltsov, D. Kivotides, N.B. Kopnin and M. Krusius for
discussions.  This work was supported by
ESF COSLAB Programme and by the Russian Foundations for
Fundamental Research.

\end{document}